\documentstyle[12pt]{article}

\newcommand{\be}{\begin{equation}}   \newcommand{\ee}{\end{equation}}
\newcommand{\bear}{\begin{eqnarray}}
\newcommand{\eear}{\end{eqnarray}}
\newcommand{\ba}{\begin{array}}      \newcommand{\ea}{\end{array}}

\newcommand{\gae}{\begin{array}{c}\,\sim\vspace{-21pt}\\> \end{array}}
\begin{document}

\pagestyle{empty}
\begin{titlepage}
\def\thepage {}        

\title{\bf Electroweak Symmetry Breaking as \\ [2mm] 
a Consequence of Compact Dimensions \\ [1cm]}

\author{\bf Bogdan A.~Dobrescu \\
\\
{\small {\it Fermi National Accelerator Laboratory}}\\
{\small {\it P.O. Box 500, Batavia, IL 60510, USA \thanks{e-mail
  address: bdob@fnal.gov} }}\\ }

\date{ }
\maketitle

   \vspace*{-10cm}
\noindent
\makebox[10.7cm][l]{December 13, 1998} FERMILAB-PUB-98/396-T \\ [1mm]
\makebox[12.8cm][l]{Revised: July 12, 1999}  hep-ph/9812349\\ 

 \vspace*{12.5cm}

\baselineskip=18pt

\begin{abstract}
   {\normalsize
It has been shown recently that the Higgs doublet may be composite,
with the left-handed top-bottom doublet and a new right-handed 
anti-quark as constituents bound by some four-quark 
operators with non-perturbative coefficients.
I show that these operators are naturally induced 
if there are extra spatial dimensions with a compactification scale 
in the multi-TeV range.
The Higgs compositeness is due mainly to the Kaluza-Klein modes of the 
gluons, while flavor symmetry breaking may be provided by various fields
propagating in the compact dimensions.
I comment briefly on the embedding of this scenario in string theory.
}

\end{abstract}

\vfill
\end{titlepage}

\baselineskip=18pt
\pagestyle{plain}
\setcounter{page}{1}


Recently it has been shown that a bound state formed 
of the left-handed top-bottom quark doublet 
and a new right-handed anti-quark can play viably the role 
of the Standard Model (SM) Higgs doublet \cite{dhseesaw, eff}.
The only ingredients necessary in this framework are some 
four-quark interactions, suppressed by a multi-TeV scale $M_1$. 
One is then led to ask what is the origin of these non-renormalizable 
interactions. The traditional answer is that they are produced by
the exchange of some heavy gauge bosons associated with the breaking 
of a larger gauge group down to the QCD group. The simplest choice of this
sort is the topcolor \cite{topcolor} scheme:
$SU(N_c)_1 \times SU(N_c)_2 \rightarrow SU(N_c)_{\rm C}$, where $N_c = 3$, 
the third generation quarks are charged under
$SU(N_c)_1$, and the first two generations are charged under
$SU(N_c)_2$.
This choice has several nice features. For example, the asymptotic freedom 
of the topcolor gauge group allows the 
heavy gauge bosons to be strongly coupled at the scale $M_1$
without problems with a Landau pole. Also, it is technically convenient
because it allows the use of the large-$N_c$ expansion.

The drawback is that one has to include additional structures in order to 
spontaneously break topcolor. Moreover,
this embedding of the $SU(N_c)_{\rm C}$ color group may be seen as 
artificial. The embedding can be improved by allowing 
all quarks to transform only under $SU(N_c)_1$ 
\cite{coloron} which is 
then embedded in a gauged flavor or family symmetry 
\cite{family, mirror}.
However, such an extension requires a more complicated 
topcolor breaking sector.

It is therefore legitimate to ask whether
one can avoid the extension of the gauge group while 
inducing the desired four-quark operators and retaining the nice 
features of topcolor ?
In this letter I point out that the required four-quark operators 
are naturally induced if the SM gauge fields 
propagate not only in the usual four-dimensional Minkowski space-time
but also in a compact space.

The first step in showing this is to recall that the existence of 
the extra dimensions is manifested within the
4-dimensional space-time through a tower of Kaluza-Klein (KK) modes 
associated with each of the fields propagating in the compact
manifold (see {\it e.g.}, \cite{kkstates, ddg}).
The KK modes of a massless gauge boson 
have masses between $M_1 = R^{-1}_{\rm max}$,
where $R_{\rm max}$ is the largest compactification radius,
and the fundamental scale associated with quantum
gravity, $M_*$.
Assuming for simplicity that the fermions propagate at the 
fixed points of an orbifold, so that they have only zero-modes, it 
follows that the couplings of the KK modes of the gauge bosons to the 
quarks and leptons are identical (up to an overall normalization) 
with those of the SM gauge bosons.
Thus, the KK excitations of the gluons give rise
in the low energy theory to flavor universal four-quark operators:
\be
{\cal L}_{\rm eff}^c =
- \frac{c g_s^2(M_1)}{2 M_1^2} 
\left(\sum_{q} \overline{q} \gamma_\mu T^a q \right)^2 ~,
\label{ops1}
\ee
where $q$ are all the quarks, $T^a$ are the $SU(N_c)_{\rm C}$
generators, 
and $g_s$ is the QCD gauge coupling.
The dimensionless coefficient $c > 0$ sums the contributions of all
gluonic KK modes, so that it depends on the number of 
extra dimensions and on the compactification radii.
Higher dimensional operators are also induced in the low energy theory,
but the contribution from a gluonic KK mode
of mass $M_n$ is suppressed compared to the contribution to $c$
by powers of $M_n/M_1$. Therefore, the effects of the higher dimensional 
operators may be neglected.
By contrast, in topcolor models the higher dimensional 
quark operators induced by the heavy gauge bosons are usually 
ignored for convenience; this procedure may be physically reasonable 
but so far has not been mathematically motivated. 

The contact interaction ${\cal L}_{\rm eff}^c$ is attractive in the scalar 
channel, so that spinless $\overline{q}_L q_R$ bound states form.
Their properties can be studied using an effective potential formalism.
In the large-$N_c$ limit only the left-right current-current part of 
${\cal L}_{\rm eff}^c$ contributes to the effective potential for the 
composite scalars. 
Note that a Fierz transformation of this current-current interaction 
gives the well known Nambu--Jona-Lasinio (NJL) interaction. Hence, the 
large-$N_c$ limit and the NJL model are equivalent approximations
of the gluonic KK dynamics.  For $c$ larger than a critical value, 
the composite scalars acquire electroweak asymmetric VEVs \cite{bhl}.
Ignoring the renormalization group evolution of $c$ above the scale $M_1$, 
one can find the critical value to leading order in $1/N_c$:
\be
c_{\rm crit} = \frac{2\pi}{N_c \alpha_s(M_1)} ~,
\ee
where $\alpha_s = g^2_s/4\pi$.
The important feature here is that the chiral phase transition 
is second order as $c$ is varied. This property is expected to remain
true beyond the large-$N_c$ approximation \cite{transition}.
In the absence of excessive fine-tuning, {\it i.e.} if $c$ is not very 
close to $c_{\rm crit}$, the electroweak scale of 246 GeV 
indicates that $M_1$ is in the multi-TeV range, or smaller.

The next step is to decide whether the
four-quark operators induced by the gluonic KK modes can 
be super-critical. In principle the answer is positive, because
the higher-dimensional gauge coupling is dimensionfull 
such that the strength of the gauge interactions increases rapidly 
above $M_1$ \cite{tcdim}.

It would be useful to find the 
dimensionality and topology of the compact manifold necessary for
electroweak symmetry breaking (EWSB).
For simplicity, consider a $\delta$-dimensional torus or orbifold with 
radii $R_l$, $l = 1,..., \delta$. The 
spectrum of the KK excitations of a massless field
is given by
\be
M_{n_1, ..., n_\delta}^2 = \sum_{l = 1}^{\delta}\frac{n_l^2}{R_l^2} ~,
\ee
where $n_l$ are integers (KK excitation numbers). In what follows, 
the KK mass levels will be denoted by $M_n$ with $n$ integer
($1 \le n \le n_{\rm max}$ where $M_{n_{\rm max}} \approx M_*$),
and their degeneracy by $D_n$. 
To compute the coefficient of the four-quark operator, one would
need to integrate out the $D_{n_{\rm max}}$ modes of mass 
$M_{n_{\rm max}}$, then to use the renormalization group evolution 
for $c$ from $M_{n_{\rm max}}$ 
down to $M_{n_{\rm max} - 1 }$, and to repeat these steps for each KK 
mass threshold until the lightest states are integrated out. 
This would give the coefficient $c$ at the scale $M_1$ as a function of 
$M_{*}/R_l$ and $g_s(M_1)$.

Fortunately it is not necessary to perform this computation in order to 
show that there are 
compact manifolds which induce EWSB.
Moreover, a continuity argument shows that $c$ may be tuned 
close to the critical value.
To see this, note that the contribution from any gluonic KK state
to $c$ is positive, so that truncating the tower of states at some 
$n_{\rm tr} < n_{\rm max}$ gives $c(n_{\rm tr}) < c(n_{\rm max}) = c$. 
Furthermore, the running of $c$ may be ignored if 
$M_{n_{\rm tr}}$ is sufficiently close to $M_1$.
For example, if the $\delta$ extra dimensions have the same
compactification radius, $R$, then 
the truncation of the KK tower at $M_{n_{\rm tr}} = 2/R$ yields
\be
c(n_{\rm tr} = 4) = \sum_{n = 1}^{4} \frac{D_n}{n}  ~,
\ee
where the degeneracy is given for $n \le 7$ by 
\be
D_n = 2^n \, \delta ! \, \left[ \frac{\theta(\delta - n)}{n!\, (\delta - n)!}
+ \frac{\theta(n - 4)\theta(\delta - n +3)}
{8 \, (n - 4)! \, (\delta - n +3)!} \right] ~,
\ee
with the step function $\theta(x \ge 0) = 1$. 
For $\delta = 4$, and $M_1$ in the multi-TeV range [where 
$\alpha_s(M_1) \approx 0.08 - 0.06$, corresponding to
$c_{\rm crit} \approx 26 - 35$],  
\be
c > c(4) \approx 35.7 > c_{\rm crit} ~,
\ee
which shows that four extra dimensions are sufficient for 
the composite Higgs doublets to acquire VEVs.
As a corollary, the SM is not viable if 
there are four or more extra dimensions with a compactification scale
below $M_*/2$, because the quarks 
would acquire dynamical masses of the order of the 
compactification scale.

Consider now the case $\delta = 1$. Ignoring the running of $c$,
the sum over the contributions from all KK states yields
\be
c = \sum_{n_1 = 1}^{n_{\rm max}} \frac{2}{n_1^2} < 
\frac{\pi^2}{3} < c_{\rm crit} ~.
\ee
Even if the contributions from the running of $c$ happen to be 
positive, it seems unlikely that they 
amount to the factor of order 10 required to drive $c$ over the 
critical value.
Thus, if there is only one compact dimension, it is fair to 
expect that the electroweak symmetry remains unbroken.

Coming back to the super-critical $\delta = 4$ case, one can imagine 
decreasing continuously three of the four radii while keeping $M_*$ fixed. 
When these radii reach the value $1/M_*$, the case $\delta = 1$ is recovered.
As a result, the coefficient $c$ of the four-quark operator (\ref{ops1})
decreases continuously (ignoring the fact that 
the number of gluonic KK modes below $M_*$, $N_{\rm KK}$, is finite) 
from the super-critical case,
$R_l = R, \; l=1,...,4$, down to the sub-critical case,
$R_1 = R, \; R_2 = R_3 = R_4 = 1/M_*$.
In the large $N_c$ limit, this decrease in the strength of the four-quark
coupling is associated with a continuous change of the mass-squared of the 
composite $\overline{q}_L q_R$ scalars, from a negative value in the
super-critical case to a positive value in the sub-critical case.
Assuming that the results obtained in the large $N_c$ limit 
are valid for $N_c = 3$, 
it follows that the decrease of some of the radii for fixed $M_*$
induces a second order transition from the 
phase in which the electroweak symmetry is broken to the unbroken phase.
The boundary of the phase transition corresponds to a set of
manifolds with radii $R_1, ..., R_\delta$, where $2 \le \delta \le 4$.
A more precise estimate of $c$ might change the lower bound on $\delta$.
On the other hand, the above arguments show that there are 4-dimensional
manifolds, with a hierarchy of compactification radii [at least
one radius has to be shorter than $1/(2M_*)$], which yield $c$ 
close to the critical value.       
This result opens up the possibility of constructing realistic
models of Higgs compositeness based on compact dimensions.

Although the gluonic KK modes may induce 
EWSB, they do not provide flavor symmetry breaking.
For super-critical $c$, ${\cal L}_{\rm eff}^c $ 
would produce an $SU(N_f)$ symmetric condensate 
and an $SU(N_f)$ adjoint of Nambu--Goldstone bosons ($N_f$ is the 
number of quark flavors).
All the quarks would acquire the same dynamical mass, related to the
electroweak scale.
It is thus necessary to identify a source of flavor symmetry 
breaking.
Also, as explained in ref.~\cite{dhseesaw, eff}, at least one new quark, 
$\chi$, should be introduced such that 
a $\overline{t}_L \chi_R$ dynamical mass of order 0.5 TeV is induced,
leading to the observed $W$ and $Z$ masses.

The KK excitations of the 
hypercharge gauge boson give rise to four-fermion operators 
which are attractive for the up-type quarks and repulsive
for the down-type quarks:
\be
{\cal L}_{\rm eff}^Y = 
- \frac{c g^{\prime 2}}{M_1^2}\left(
\frac{1}{3}\overline{\psi}_L^i \gamma_\mu  \psi_L^i 
+ \frac{4}{3}\overline{u}_R^i  \gamma_\mu u_R^i 
- \frac{2}{3}\overline{d}_R^i  \gamma_\mu d_R^i 
+  \frac{4}{3} \overline{\chi} \gamma^\mu \chi 
\right)^2 ~,
\label{ops2}
\ee
where $i = 1,2, 3$ is a generational index, $\psi^i_L = (u^i ,d^i )_L$, 
and the lepton currents are not shown for simplicity. The 
vector-like quark $\chi$ transforms under the SM gauge group
in the same representation as $u^i_R$.
If the gluonic KK modes yield $c$
within about 10\% of its critical value, then the combination 
${\cal L}_{\rm eff}^c + {\cal L}_{\rm eff}^Y$ induces VEVs only 
for the Higgs fields made up of the $u, c, t$  and $\chi$ quarks.

At this stage it is necessary to introduce inter-generational
flavor symmetry breaking. It is convenient to parametrize it
using four-fermion operators of the following type:
\be
\frac{\eta_{AB} }{M_1^2} \left( \overline{A}_L B_R \right) 
\left( \overline{B}_R A_L \right) ~,
\label{flavor}
\ee
with the notation
$A_L = \psi^i_L, \chi_L $ and $B_R = u^k_R, d^i_R$, where
$k = 1, ..., 4$, and $u^4_R \equiv \chi_R$.
Unlike the four-quark operators 
induced by the gluonic KK modes, which are strongly 
coupled and give rise to deeply bound states, the  
operators (\ref{flavor}) can be treated perturbatively because 
the coefficients $\eta_{AB}$ do not need to be larger than ${\cal O}(1)$.

The origin of these operators can be found in different 
scenarios for the physics at scales above $M_1$. 
For example, they can be produced at the fundamental scale
$M_*$, where the most general gauge invariant operators are likely
to be present. Note that the 
't Hooft coupling $\alpha_s N_{KK}$
becomes non-perturbative not far above 
$M_1$, at a scale which is likely to be $M_*$. 
Such a low $M_*$ may occur in the truly strong coupling 
regime of string theory \cite{lykken}, as well as when
there are large dimensions inaccessible to the SM fields
\cite{largedim}. 
Predicting the coefficients of the flavor non-universal operators 
may require a complete theory that includes quantum gravitational effects
[this also applies to the flavor violation in eq.~(\ref{yukmas}) below]. 
Another possibility is that the flavor breaking operators are 
generated by various fields propagating in the large extra dimensions 
\cite{tcdim, flavor}. Also, if the position of the quarks in the extra
dimensions is flavor-dependent \cite{martin}, then even flavor-universal 
interactions at the scale $M_*$ would lead to non-universality in 
eq.~(\ref{flavor}).

If one of the $\eta_{\psi^i u^k}$ coefficients, chosen by convention 
to be $\eta_{\psi^3 \chi}$, is larger than the other 
eleven, then the condition
\be
\eta_{\psi^3 \chi} > 
\frac{2 \pi^2}{N_c} - c \left( g_s^2 + \frac{8}{9 N_c} g^{\prime 2}\right) >
\eta_{\psi^i u^k} ~
\ee
implies that only 
the $\overline{\chi}_R \psi^3_L$ Higgs doublet
has a negative squared mass, resulting in the hierarchy between the 
top quark and the others.

In order to accommodate the observed
masses of the $W$, $Z$ and $t$, a few other conditions must be 
satisfied \cite{dhseesaw,eff}.
First, the gauge invariant mass term 
$\mu_{\chi\chi}\overline{\chi}_L \chi_R$ 
has to be included, so that tadpole terms for the 
$\overline{\chi}_L \chi_R$ scalar are induced in the effective potential.
For this purpose, the $\mu_{\chi\chi}$ mass parameter can be significantly 
smaller than $M_1$. Such a small mass may arise 
naturally, for example from the VEV of a gauge singlet scalar which propagates
in some compact dimensions which are inaccessible to $\chi$
(a similar mechanism is used in 
ref.~\cite{neutrinos} to produce neutrino masses).
Generically, the $\mu_{\chi u^i}\overline{\chi}_L u^i_R$ mass terms
are also present.
A second condition for realizing the top condensation seesaw mechanism
\cite{dhseesaw,eff} is to have the squared masses of the 
$\overline{\chi}_R \chi_L$ and $\overline{t}_R \chi_L$ scalars
larger than the squared mass of the $\overline{\chi}_R \psi^3_L$ Higgs 
doublet, or equivalently: 
$\eta_{\psi^3 \chi} > \eta_{\chi t}, \, \eta_{\chi\chi}$.
These conditions ensure that 
the minimum of the effective potential for composite scalars 
corresponds to dynamical fermion masses only for the $t$ and $\chi$.
The $\overline{t}_L \chi_R$ mass mixing, responsible for 
EWSB, has to be of order 0.5 TeV, 
while the top mass measurements and the constraint on 
custodial symmetry violation require the fermion mass eigenvalues 
to be $m_t \approx 175$ GeV and 
$m_\chi \gae 3$ TeV. In the absence of excessive fine-tuning,
$m_\chi \sim {\cal O}(5)$ TeV corresponds to a compactification scale
$M_1$ of order 10 -- 50 TeV.

The gluonic KK excitations are responsible for the 
existence of $N_f^2$ composite complex scalars, each of the left-handed
quark flavors binding to each of the right-handed ones.
Their mass degeneracy is lifted by the flavor non-universal four-fermion
operators (\ref{flavor}), but generically most of the physical states
have masses of order $m_\chi$ \cite{eff}.
However, the composite scalar spectrum includes 
the longitudinal $W$ and $Z$, and a neutral Higgs boson which
is always lighter than 1 TeV, and may be as light as ${\cal O}(100)$ GeV
if the vacuum is close to the boundary of a second order phase transition 
\cite{eff}. 
In the decoupling limit, where $m_\chi \rightarrow \infty$,
the low energy theory is precisely the SM, with the possible
addition of other composite states which may be light due to the vicinity 
of a phase transition.

So far, the electroweak symmetry is broken correctly, $m_t$
is accommodated, and the $\chi$ quark has a mass of at least
a few TeV. It remains to produce the masses and mixings of the other quarks 
and leptons. For this reason, consider the following four-fermion operators,
assumed to be produced by physics above the compactification scale:
\be
\frac{1}{M_1^2} \left( \overline{\chi}_R \psi_L^3\right) 
\left[ \xi_{\psi^j u^k} \left( \overline{\psi}_L^j u_R^k\right) +
\xi_{\psi^j d^k} \left( \overline{\psi}_L^j i\sigma_2 d_R^k\right) +
\xi_{l^j \nu^k} \left( \overline{l}_L^j \nu_R^k\right) +
\xi_{l^j e^k} \left( \overline{l}_L^j i\sigma_2 e_R^k\right) 
\right] ~,
\label{yukmas}
\ee
where $l_L^j$, $\nu_R^j$ and $e_R^j$ are the SM
lepton fields.
These operators lead through the renormalization group evolution to
SM Yukawa couplings (proportional with the dimensionless coefficients $\xi$)
of the $\overline{\chi}_R \psi_L^3$ composite 
Higgs doublet to the fermions. 
Another effect of the operators (\ref{yukmas}) is to mix
the $\overline{\chi}_R \psi_L^3$ 
Higgs doublet with the other $\overline{q}_R^i \psi_L^j$ composite
scalars. 

The flavor breaking operators (\ref{flavor}) and (\ref{yukmas})
do not produce large  flavor-changing neutral current (FCNC) effects 
beyond those in the SM. 
It remains to be shown however that these
operators can be induced by some high energy physics without 
being accompanied by flavor-changing operators with large 
neutral-current effects. Another contribution to the FCNCs
comes from the composite scalars, which have flavor breaking
couplings to the quark mass-eigenstates. These are suppressed by 
Kobayashi-Maskawa elements, and they are not worrisome if the
scalars are sufficiently heavy. It is beyond the scope of this letter
to derive the constraints from FCNCs, but it is worth reiterating 
that the extra dimensions open up new ways of dealing with flavor.
In addition, the low energy theory is flexible: for example 
the four-quark operators may be kept flavor universal by allowing
flavor violation to arise from the mass terms of an extended vector-like 
quark sector \cite{simmons, family}.

Placing the fermions at orbifold fixed points is convenient because 
all the gluonic KK modes contribute in this case at tree level to
the four-quark operators.
However, if at least one right-handed up-type quark propagates in 
compact dimensions, its KK modes could play the role of the vector-like quark, 
$\chi$, required to account for the bulk of EWSB. 
This possibility requires further 
study because the couplings of the quark and gluonic KK modes 
are restricted by momentum conservation in the compact dimensions.

It is instructive to see how the scenario presented here may arise
from string theory or M theory.
The most convenient setting for studying large extra dimensions
is within Type I string theory \cite{ddg, lykken, typeI}. 
The closed string sector gives rise to the  graviton and other 
neutral modes which propagate in the bulk of the 9+1 dimensional 
space-time.
The open string sector gives rise to the gauge fields and the 
charged matter, which are restricted to propagate on a D9-brane
or a D5-brane. Using T-duality transformations, one may 
obtain Type I$^\prime$ string theories containing D$p$-branes
with $p \le 9$. 
A $(\delta + 3)$-brane, with $2 \le \delta \le 4$, is necessary for 
containing the 3+1 dimensional 
flat space-time plus the $\delta$ compact dimensions that lead to 
a composite Higgs sector. 
Some of the $\delta$ compactification scales are expected to be of
order $M_1 \sim 10 - 50$ TeV, while other must be slightly higher,
in order to allow the coefficients of the four-quark operators to be
close to criticality. 

Due to the presence of the KK modes of the SM 
gauge bosons, the gauge couplings tend to unify at a scale higher 
by at most one order of magnitude than the compactification scale 
\cite{ddg}. The running of the gauge couplings in the 
model discussed here is different than in the supersymmetric 
SM because below $M_1$
there are no superpartners, and there is a potentially 
complicated composite Higgs sector.
Nevertheless, the gauge coupling unification 
may be realized in various ways, due to the possible existence of
additional states below or above $M_1$.
Alternately, the $SU(3)_C \times SU(2)_W \times U(1)_Y$ 
gauge groups may come from different types of branes, so that 
the gauge couplings need not unify precisely.

It is natural to identify the unification scale with the 
string scale, so that $M_s \sim {\cal O}(100)$ TeV, 
which also corresponds to the scale where $\alpha_s N_{KK} \sim 1$. 
This prediction is a 
consequence of the fit to the electroweak scale and top-quark mass
within the top seesaw theory of Higgs compositeness.
The low $M_s$ is associated with large extra dimensions accessible only to 
gravity \cite{largedim}. If these are identified with the 
$6 - \delta$ dimensions orthogonal to the $(\delta + 3)$-brane, then their 
compactification radius is given by
\be
r^\prime \sim \frac{1}{M_s}
\left[ \alpha_s(M_s) \left(M_s^\delta R_1 ... R_\delta\right)^{\! 1/2}
\frac{M_{\rm Planck}}{M_s} 
\right]^{\frac{2}{(6 - \delta)}} ~.
\ee
Using the value 
$\alpha_s(M_s) \sim 1/50$ for the gauge coupling at the string scale,
$M_{\rm Planck} \sim 10^{19}$ GeV, and $R_1 ... R_\delta \sim 
(30 \; {\rm TeV})^\delta$ 
gives $r^\prime \sim 10^{-6}$ cm for $\delta = 4$, and
$r^\prime \sim 10^{-10}$ cm for $\delta = 3$.
Due to the low $M_s$, there may be stringy effects that change 
the physics at the scale $M_1$, including  
the coefficients of operators induced by gluonic KK modes. 
Whether the picture described here changes qualitatively 
remains an open question, together with other issues in
the context of D-branes, such as how to construct stable 
non-supersymmetric brane configurations, or what branes
correspond to the three generations of chiral fermions \cite{brane}. 

In conclusion, the top condensation seesaw mechanism is a compelling 
scenario for EWSB. It is remarkable that 
viable models involving this mechanism do not require an extension
of the SM gauge group, provided there are a few 
compact dimensions accessible to the gluons. If the future collider 
experiments will probe the composite Higgs sector, 
we will be able to test the existence of the
KK modes of the gluons, and consequently the structure of the 
space-time. 

\vspace{.1mm}
\noindent
{\it Acknowledgements:} I thank N.~Arkani-Hamed,
W.~Bardeen, G.~Burdman, H.-C.~Cheng, S.~Chivukula, 
N.~Evans, C.~Hill, and J.~Lykken for helpful discussions and 
comments.

\vspace{-2mm}

\vfil

\begin{thebibliography}{99}
\frenchspacing

\bibitem{dhseesaw}
 B.~A.~Dobrescu and C.~T.~Hill, Phys.~Rev.~Lett.~{\bf 81} (1998) 2634 .
\bibitem{eff} R.~S.~Chivukula, B.~A.~Dobrescu, H.~Georgi, and C.~T.~Hill,
 Phys.~Rev.~{\bf D59} (1999) 075003.
\bibitem{topcolor} C.~T.~Hill, Phys. Lett.~{\bf B266} (1991) 419.
\bibitem{coloron} R.~S.~Chivukula, A.~G.~Cohen, and E.~H.~Simmons,
 Phys.~Lett.~{\bf B380} (1996) 92;
 E.~H.~Simmons, Phys.~Rev.~{\bf D 55} (1997) 1678;
 M.~B.~Popovic and  E.~H.~Simmons, Phys.~Rev.~{\bf D58} (1998) 095007. 
\bibitem{family} G.~Burdman and N.~Evans, Phys. Rev.~{\bf D59} (1999) 115005.
\bibitem{mirror} M.~Lindner and G.~Triantaphyllou,
	Phys. Lett.~{\bf B430} (1998) 303; G.~Triantaphyllou, hep-ph/9811250.
\bibitem{kkstates} I.~Antoniadis, Phys.~Lett.~{\bf B246} (1990) 377; 
	V.~A.~Kostelecky and S.~Samuel, Phys. Lett.~{\bf B270} (1991) 21:
I.~Antoniadis and K.~Benakli, Phys.~Lett.~{\bf B326} (1994) 69;
I.~Antoniadis, K.~Benakli and M.~Quiros, Phys.~Lett.~{\bf B331} (1994) 313.
\bibitem{ddg} K.~R.~Dienes, E.~Dudas, and T.~Ghergheta, 
	Phys. Lett. {\bf B436} (1998) 55; Nucl. Phys. {\bf B537} (1999) 47.
\bibitem{bhl} W.A.~Bardeen, C.T.~Hill and M.~Lindner, Phys. Rev. {\bf
	D41}, 1647 (1990).
\bibitem{transition} R.~S.~Chivukula and H.~Georgi, Phys.~Rev.~{\bf D58}, 
	075004 (1998). 
\bibitem{tcdim} N.~Arkani-Hamed and S.~Dimopoulos, hep-ph/9811353.
\bibitem{lykken} J.~Lykken, Phys.~Rev.~{\bf D54}, 3693 (1996).
\bibitem{largedim} N.~Arkani-Hamed, S.~Dimopoulos and G.~Dvali, 
	Phys.~Lett.~{\bf B429}, 263 (1998).
\bibitem{flavor} Z.~Berezhiani and G.~Dvali, hep-ph/9811378.
\bibitem{martin} N.~Arkani-Hamed and M.~Schmaltz, hep-ph/9903417. 
\bibitem{simmons} E.~H.~Simmons, Nucl.~Phys.~{\bf B324}, 315 (1989).
\bibitem{neutrinos} K.~R.~Dienes, E.~Dudas, T.~Ghergheta, hep-ph/9811428; 
N.~Arkani-Hamed, S.~Dimopoulos, G.~Dvali and J.~March-Russell, hep-ph/9811448.
\bibitem{typeI} 
I.~Antoniadis, N.~Arkani-Hamed, S.~Dimopoulos and G.~Dvali, 
	Phys.~Lett.~{\bf B436}, 257 (1998); 
G~Shiu and S.-H.~H.~Tye, Phys.~Rev.~{\bf D58}, 106007 (1998);
Z.~Kakushadze and S.-H.~H.~Tye, hep-th/9809147. 
\bibitem{brane} J.~Lykken, E.~Poppitz, S.~P.~Trivedi, 
	Nucl.~Phys.~{\bf B543} (1999) 105.

\end{thebibliography}
\end{document}